\documentclass[prl,letterpaper,twocolumn,tightenlines,superscriptaddress,
showpacs,preprintnumbers,nofootinbib]{revtex4}
\usepackage{amsfonts}
\usepackage{graphicx}

\usepackage{amssymb}                                                          
\usepackage{amsmath}

\newcommand{\hfn}{{\sc hfn}}
\newcommand{\hfa}{{\sc hfa}}
\newcommand{\dar}{{\sc dar}}
\newcommand{\hfndar}{{\sc dar+hfn}}
\newcommand{\hfnhfa}{{\sc hfa+hfn}}

\begin{document}
\pagestyle{plain}

\preprint{VPI-IPNAS-10-09}

\title{A new approach to anti-neutrino running in long baseline neutrino oscillation experiments}

\author{Sanjib K. Agarwalla}
\email{sanjib@vt.edu}

\author{Patrick Huber}
\email{pahuber@vt.edu}

\author{Jonathan M. Link}
\email{jonathan.link@vt.edu}

\author{Debabrata Mohapatra}
\email{dmohapat@vt.edu}

 \affiliation{Department of Physics, Virginia
  Tech, Blacksburg, VA 24060, USA}

\date{\today}

\begin{abstract}
  We study the possibility to replace the anti-neutrino run of a long
  baseline neutrino oscillation experiment, with anti-neutrinos from
  muon decay at rest. The low energy of these neutrinos allows the use
  of inverse beta decay for detection in a Gadolinium-doped water
  Cerenkov detector. We show that this approach yields a factor of
  five times larger anti-neutrino event sample. The resulting
  discovery reaches in $\theta_{13}$, the mass hierarchy and leptonic
  CP violation are compared with those from a conventional superbeam
  experiment with combined neutrino and anti-neutrino running. We find
  that this approach yields a greatly improved reach for CP violation
  and $\theta_{13}$ while leaving the ability to measure the mass
  hierarchy intact.
\end{abstract}
\pacs{14.60.Lm, 14.60.Pq, 29.20.dg, 29.40.Ka} 

\maketitle

The fact that our Universe is made out of matter and not anti-matter
constitutes one of the great open questions in physics. The conditions
required to obtain a baryon asymmetry from a CP symmetric initial
state were identified by Sakharov more than 50 years
ago~\cite{sakharov}. They are deviation from thermal equilibrium,
baryon number violation and CP violation. Of these three ingredients,
CP violation has proven to be the most elusive. While some small
amount of CP violation exists in quark
mixing~\cite{Christenson:1964fg,Barberio:2008fa}, it is not sufficient
to explain the baryon asymmetry of the Universe~\cite{Riotto:1999yt}.
Therefore, the quest for new sources of CP violation continues to be a
guiding theme for much of particle physics. Leptonic CP violation has
been proposed as a mechanism to explain the baryon asymmetry of the
Universe, so called leptogenesis~\cite{Fukugita:1986hr}, and with the
discovery of neutrino oscillation~\cite{Maltoni:2004ei} leptogenesis
has become a very plausible scenario. The discovery of CP violation in
the lepton sector would constitute a smoking gun for this mechanism.
In addition the mere existence of neutrino oscillations implies a
neutrino mixing matrix in the same fashion as the CKM matrix in the
quark sector, and with three generations there will be at least one CP
phase.

Studying CP violation in neutrino oscillation, requires the use of
appearance channels, when the neutrino changes flavor from production
to detection. Since $\nu_\tau$ are very hard to detect and to produce,
naturally, all proposed methods focus on using transitions between
$\nu_e\leftrightarrow\nu_\mu$ and
$\bar\nu_e\leftrightarrow\bar\nu_\mu$. There are various proposals for
novel techniques to produce neutrino and anti-neutrino beams, either
from the decays of muons (neutrino factories), or from the
$\beta$-decay of short-lived, artificially produced radioactive nuclei
($\beta$-beams), see {\it e.g.}~\cite{Bandyopadhyay:2007kx}. While
these concepts may ultimately provide the best sensitivity to CP
violation, in the near term conventional neutrino and anti-neutrino
beams based on the decay of horn focused pions are the option of
choice (superbeams). All beam based approaches to neutrino oscillation
are performed with neutrinos in the energy range from a few hundred to
many thousand MeV\,. In this energy range charged current reactions
provide the dominant detection mechanism.  At the atmospheric mass
squared difference these energies imply a source detector distance
from $100\,\mathrm{km}$ up to several $1,000\,\mathrm{km}$, hence
these experiments are known as long baseline experiments.

\begin{figure*}
  \includegraphics[width=\textwidth]{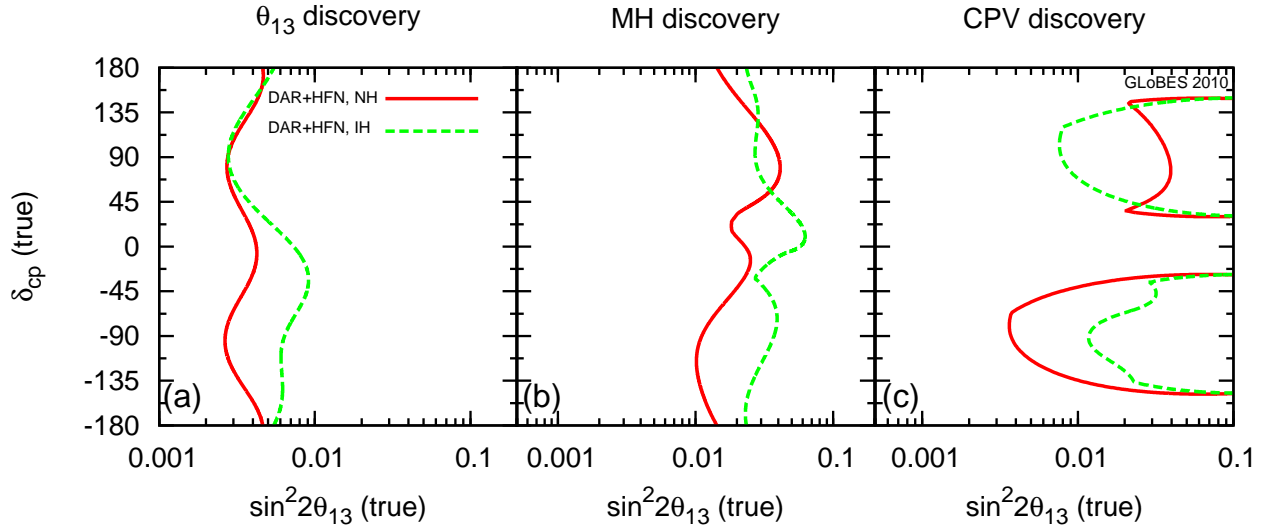}
  \caption{\label{fig:sens} Discovery reaches at $3\,\sigma$
    confidence level from left to right for $\theta_{13}$, mass
    hierarchy and CP violation for 6 years of {\hfndar} for normal
    (NH) and inverted (IH) true mass hierarchy.}
\end{figure*}

In the context of superbeam experiments, a CP violation measurement
requires data from both $\nu_\mu\rightarrow\nu_e$ \emph{and}
$\bar\nu_\mu\rightarrow\bar\nu_e$. However, the anti-neutrino run
poses a number of specific challenges: the production rate for
$\pi^-$, the parent of $\bar\nu_\mu$, is lower than for $\pi^+$, the
anti-neutrino charged current cross section is lower, the background
levels are higher\footnote{This is due to the larger contamination from wrong
  sign pions.}, and the systematic errors are expected to
be larger. Overall, the event rate for anti-neutrinos is suppressed by
a factor of 2-5, depending on the anti-neutrino energy, which is
illustrated by table~\ref{tab:rates}.

On the other hand, for energies below about $100\,\mathrm{MeV}$ the
situation is quite the opposite, strongly favoring the use of
anti-neutrinos.  In this energy range the dominant process is inverse
$\beta$-decay, (IBD), $\bar\nu_e + p \to e^{+} + n$. This process
provides a very useful delayed coincidence tag between the prompt
positron and the delayed neutron capture, which is the reason why this
reaction was used by Reines and Cowan to discover the
neutrino~\cite{Reines:1956rs}. At these energies copious sources of
anti-neutrinos exist. In this letter we focus on $\bar\nu_\mu$ from
$\mu^+$ decay at rest from a stopped pion beam.

In a stopped pion source a proton beam of a few GeV energy interacts
in a target producing both $\pi^+$ and $\pi^-$. The pions then are
brought to rest in a high-Z beam stop. The $\pi^-$, and also the
daughter $\mu^-$, will capture on the high-Z nuclei. The $\pi^+$ will
produce the following cascade of decays
\begin{eqnarray*}
\pi^+ &\rightarrow &\mu^+ +\nu_\mu \\
&& \hspace{0.1cm}\raisebox{0.5em}{$\mid$}\!\negthickspace\rightarrow\ e^+ + \nu_e +\bar\nu_\mu 
\end{eqnarray*} 
resulting in $\nu_\mu$, $\bar\nu_\mu$ and $\nu_e$, but no $\bar\nu_e$.
The $\bar\nu_\mu$ can oscillate into a $\bar\nu_e$ and this
$\bar\nu_e$ can be detected and uniquely identified by IBD. This
concept has been used for instance in the LSND
experiment~\cite{Athanassopoulos:1996jb}. We will denote this
combination of anti-neutrino source and IBD detection as muon decay at
rest ({\dar}).

The central idea of this letter is to combine a horn focused high
energy $\nu_\mu$ beam, henceforth denoted as horn focused neutrino
beam ({\hfn}) with $\bar\nu_\mu$ from a {\dar} setup to study
$\theta_{13}$, the mass hierarchy and leptonic CP violation. This is
in contrast to reference~\cite{conrad}, where an entirely {\dar} based
experiment was considered to measure leptonic CP violation. In a pure
{\dar} setup there is no information on the mass hierarchy and
therefore a CP measurement is only feasible assuming that the mass
hierarchy has been determined by other means. We show that the
combination of {\hfn} and {\dar} allows one to address both the mass
hierarchy and CP violation at the same time. Throughout this letter,
we will denote this new technique as {\hfndar}.

To illustrate the strength of {\hfndar}, we will study a specific
setup, which closely resembles the Fermilab DUSEL concept for a long
baseline experiment, currently known as LBNE. Obviously, similar
considerations hold for any superbeam experiment. This setup has a total
running time of 6 years and a $100\,\mathrm{kt}$ water Cerenkov
detector.  The entire {\hfn} part is very similar to the setup
described in detail in~\cite{wbb}, specifically we take the source
detector distance to be $1300\,\mathrm{km}$ and use the same detector
performance. The beam delivers $6.2\times10^{20}$ protons on target
per year, which for $120\,\mathrm{GeV}$ protons corresponds roughly to
$700\,\mathrm{kW}$ of beam power.

The feasibility of a high intensity {\dar} source hinges on the
availability of cost effective, megawatt proton accelerators in the
GeV range. We use the same proton source parameters as in
reference~\cite{conrad} which are $4\times10^{22}$ of $\nu_e$,
$\nu_\mu$ and $\bar\nu_\mu$ per flavor per year per accelerator. We
use 4 of these accelerators with a source detector distance of
$20\,\mathrm{km}$.  We determined this distance to be the optimal
baseline for our purposes, by performing a scan of baselines in
$1\,\mathrm{km}$ increments from $4-24\,\mathrm{km}$. We also tested
that putting all sources at the same distance provides the best
performance.  In order to allow for an efficient detection of the IBD
signal, the water Cerenkov detector will need to be doped with
Gadolinium (Gd)~\cite{gadz}. Gadolinium-doping decreases the neutron
capture time and improves the neutron capture signature. Recent
experimental tests of this concept~\cite{gdsk} indicate a detection
efficiency of $67 \%$, which we will use for our calculation.  Due to
the strong $\pi^-$ absorption in the target, the $\bar\nu_e$
contamination is very small $\sim4 \times 10^{-4}$, but fully taken
into account~\cite{Athanassopoulos:1996jb}.  The non-beam backgrounds,
which mostly stem from atmospheric neutrino interactions are taken
from reference~\cite{conrad} and scaled to a 6 year run with a
$100\,\mathrm{kt}$ detector. The energy resolution for IBD events is
identical to the one for positrons and is parametrized as
$\sigma(E)=5\%\sqrt{E/\mathrm{GeV}}$~\cite{Nakahata:1998pz}.

The very different duty factors, $d$, of the {\hfn} source,
$d<10^{-4}$, and {\dar} source, $d\simeq0.1$, allow for concurrent
operation. As shown in table~\ref{tab:rates}, the event numbers from
both sources are comparable and thus the probability of finding events
from both source happening at the same time is approximately given by
the ratio of their duty factors, which is 0.001. This is equivalent to
less than 0.5 events for the duration of the experiment.
Obviously, for a superbeam it is not possible to run neutrinos and
anti-neutrinos simultaneously, since the horn\footnote{With a solenoid
  it is possible to focus both types of pions, but that leaves the
  problem of lepton charge identification in the detector.} either
focuses $\pi^+$ or $\pi^-$.

Note, that the disparate baselines reflect the different neutrino and
anti-neutrino energies and the ratio of baseline to energy ($L/E$) is
nearly the same for {\hfn} and {\dar}.  For comparison we also present
results for a pure superbeam experiment of the same total duration,
target mass and number of protons on target per year, which consists
of 3 years of {\hfn} operation followed by 3 years of horn focused
anti-neutrino ({\hfa}) operation. The performance for both periods is
taken from reference~\cite{wbb}. This setup is denoted by {\hfnhfa}.

In table~\ref{tab:rates}, we compare the total event rates for the 6
year {\hfndar} and {\hfnhfa} runs. For the oscillation parameters
chosen, {\hfndar} has twice the statistics in the neutrino mode and
five times as much statistics with a five times better signal to
background ratio in the anti-neutrino mode.

\begin{table}[h!]
\begin{tabular}{||c||c|c||c|c||}
\hline
\hline
&$\bar\nu_\mu\rightarrow\bar\nu_e$&bgn&$\nu_\mu\rightarrow\nu_e$&bgn\\
\hline
\hline
{\hfndar} &398 &73 &511 &143\\
\hline
{\hfnhfa}&77  &53 &255 &71\\
\hline
\hline
\end{tabular}

\caption{\label{tab:rates} Comparison of the signal and background event 
  rates of 6 years running of {\hfndar} and {\hfnhfa}. Note, that for  {\hfndar} this is 6 years of simultaneous running of neutrino and anti-neutrinos, whereas for {\hfnhfa} this is 3 years each, run consecutively. Oscillation parameters are $\sin^22\theta_{13}=0.1$ and $\delta_\mathrm{CP}=0$ and 
  normal hierarchy.}
\end{table}

\begin{figure*}[t]
  \includegraphics[width=\textwidth]{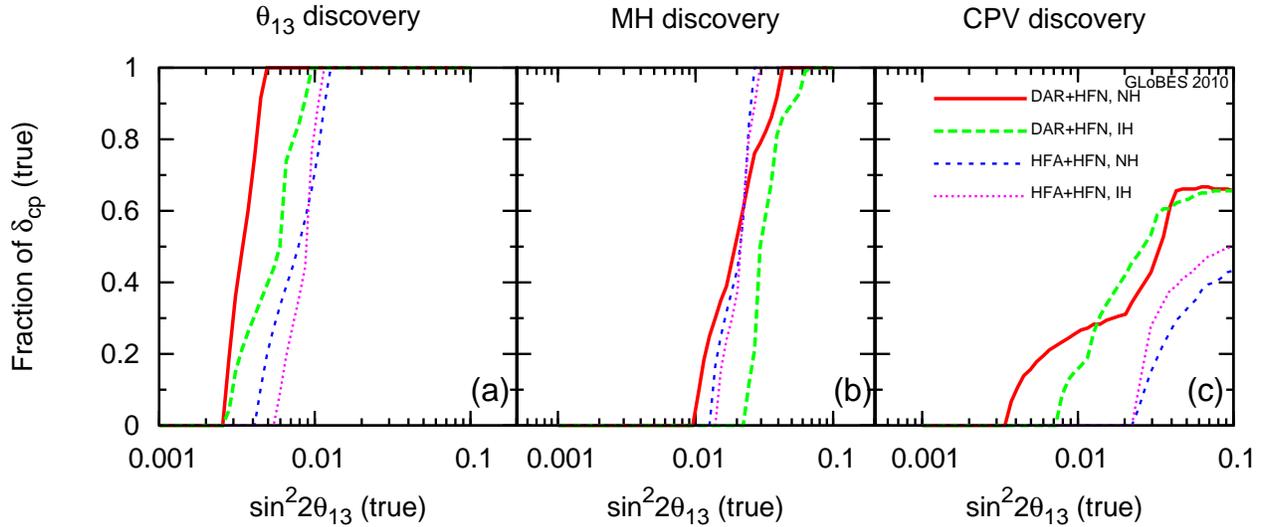}
  \caption{\label{fig:cpfrac} CP fractions for which a discovery at
    $3\,\sigma$ confidence level is possible as function of
    $\sin^22\theta_{13}$. From left to right for $\theta_{13}$, mass
    hierarchy and CP violation. The different lines are for normal
    (NH) and inverted (IH) true mass hierarchies and for {\hfndar} and
    {\hfnhfa}, respectively.}
\end{figure*}

For our statistical analysis we use the techniques outlined
in~\cite{Huber:2002mx}. We bin our {\dar} data into bins of
$1\,\mathrm{MeV}$ and the {\hfn}/{\hfa} data into bins of
$125\,\mathrm{MeV}$. We include a $5\%$ systematic on the total number
of signal events and a (uncorrelated) $5\%$ systematic on the total
number of background events.  All physics sensitivities have been
computed using {\sf GLoBES}~\cite{globes} and are shown at $3\,\sigma$
confidence level.  The oscillation parameters are
$\sin^2\theta_{12}=0.3$, $\sin^2\theta_{23}=0.5$, $\Delta
m^2_{31}=2.4\times10^{-3}\,\mathrm{eV}^2$, $\Delta
m^2_{21}=7.9\times10^{-5}\,\mathrm{eV}^2$ and we include a 10\% error
on the atmospheric parameters, a 5\% error on the solar parameters and
a 2\% matter density uncertainty. In the fit the mass hierarchy is
free.

In figure~\ref{fig:sens} we show the resulting discovery reaches for
$\theta_{13}$, the mass hierarchy and CP violation for {\hfndar} as a
function of the true values of $\sin^22\theta_{13}$ and
$\delta_\mathrm{CP}$ for both the normal and inverted true hierarchy
as labeled in the legend.  Due to the asymmetric nature of this
experiment and the fact that matter effects are relevant only for the
{\hfn} run, there is a pronounced effect from changing the input
(true) mass hierarchy.  However, in both cases the synergy between the
{\hfn} and {\dar} is apparent.

In figure~\ref{fig:cpfrac}, we compare the results from {\hfndar} with
{\hfnhfa}. The reaches are given as a fraction of $\delta_\mathrm{CP}$
and as a function of the true value of $\sin^22\theta_{13}$. In panel
(a), we show the results for the discovery of the $\theta_{13}$ and
find that {\hfndar} outperforms the superbeam experiment {\hfnhfa} for
all CP phases and both hierarchies by roughly a factor two. The
discovery reach for the mass hierarchy is shown in panel (b) and here,
we see that for some values of the CP phase, in particular for
inverted mass hierarchy, the reach is somewhat smaller for {\hfndar}.
If at the end of the {\hfndar} run, the mass hierarchy has not been
discovered adding a {\hfa} run may be required. Finally, in panel (c)
the discovery reach for CP violation is shown. {\hfndar} improves the
reach for small $\theta_{13}$ by a factor between 3 and 10 depending
on the mass hierarchy and by about 75\% for large $\theta_{13}$.

Summarizing, we have demonstrated that a combination of low energy
$\bar\nu_\mu$ from muon decay at rest with high energy $\nu_\mu$ from
a superbeam aimed at the same Gadolinium-doped water Cerenkov
detector yields a moderately improved reach for $\theta_{13}$ and a
vastly improved discovery reach for CP violation while only marginally
affecting the mass hierarchy sensitivity. These improvements are a
direct result of combining an optimized neutrino with an optimized
anti-neutrino run. The practicality of this proposal, however, depends
critically on the feasibility of low-cost, high-intensity stopped pion
sources.

We would like to acknowledge useful discussions with L.~Piilonen and
E.~Cristensen.  This work has been supported by the U.S. Department of
Energy under award numbers \protect{DE-SC0003915} and
\protect{DE-FG02-92ER40709}.

\end{document}